\documentclass[cits]{PoS}

\title{Spectra of heavy-light and heavy-heavy mesons containing 
charm quarks, including higher spin states for $N_f = 2+1$}

\ShortTitle{Spectra of heavy-light and heavy-heavy mesons}

\author{G.~Bali$^a$, S.~Collins$^a$, S.~D\"urr$^{b,c}$, Z.~Fodor$^{b,c,d}$, R.~Horsley$^e$, 
        C.~Hoelbling$^b$, S.D.~Katz$^{b,d}$, I.~Kanamori$^a$, S.~Krieg$^{b,c}$, 
        T.~Kurth$^b$, L.~Lellouch$^f$, T.~Lippert$^{b,c}$, C.~McNeile$^b$, 
        Y.~Nakamura$^g$, D.~Pleiter$^{a,c}$, \speaker{P.~P\'erez-Rubio}$^a$, 
        P.~Rakow$^h$, A.~Sch\"afer$^a$, 
        G.~Schierholz$^i$, K.K.~Szabo$^b$, F.~Winter$^e$, J.~Zanotti$^e$\\
\llap{$^a$} Institut für Theoretische Physik, Universit\"at Regensburg, 
            D-93040 Regensburg, Germany.\\
\llap{$^b$} Bergische Universit\"at Wuppertal, 
            Gaussstr. 20, D-42119, Germany.\\
\llap{$^c$} J\"ulich Supercomputing Centre, Forschungszentrum J\"ulich,
            D-52425 J\"ulich, Germany.\\
\llap{$^d$} Institute for Theoretical Physics, E\"otv\"os University,
            H-1117 Budapest, Hungary.\\
\llap{$^e$} School of Physics, University of Edinburgh, 
            Edinburgh EH9 3JZ, UK.\\
\llap{$^f$} Centre de Physique Th\'eorique\footnote{CPT is research unit
            UMR 6207 of the CNRS and of the universities Aix-Marseille I, 
            Aix-Marseille II and Sud Toulon-Var, and is 
            affiliated with the FRUMAM.}, Case 907, Campus de Luminy, 
            F-13288 Marseille, France.\\
\llap{$^g$} RIKEN Advanced Institute for Computational Science, 
            Kobe, Hyogo 650-0047, Japan.\\
\llap{$^h$} Theoretical Physics Division, Department of Mathematical 
            Sciences, University of Liverpool, Liverpool L69 3BX, UK\\
\llap{$^i$}  Deutsches Elektronen-Synchrotron DESY, 22603 Hamburg, Germany.

E-mail: \email{Paula.Perez-Rubio@physik.uni-regensburg.de }}

\abstract{We study the spectra of heavy-light and heavy-heavy mesons
  containing charm quarks, including higher spin states. We use two
  sets of $N_f=2+1$ gauge configurations, one set from QCDSF using the SLiNC action,
  and the other configurations from the Budapest-Marseille-Wuppertal
  collaboration, using the HEX smeared clover action.  To extract
  information about the excited states, we choose a suitable basis of
  operators to implement the variational method.}

\FullConference{The XXIX International Symposium on Lattice Field Theory, 
                Lattice2011\\
		July 11-16, 2011\\
		The Village at Squaw Valley, California, USA}

\begin{document}

\section{Introduction}

Charm physics has undergone a renaissance in recent years. 
The spectroscopy of hadrons
containing charm quarks has been given prominence through the
discovery of several new narrow charmonium resonances close to the
$D\overline{D}$ thresholds and new narrow $D_s$ mesons close to the $DK$
thresholds~(for reviews see
refs.~\cite{TheHeavyFlavorAveragingGroup:2010qj,Brambilla:2010cs}). These
states are likely to require the extension of non-relativistic quark
models to include the 4-quark sector. A wealth of results is expected
over the next few years from completed and existing experiments, for
example BaBar, Belle, BES-III and the LHC experiments, and the
future PANDA experiment at the FAIR facility. 

Lattice calculations of the spectroscopy of hadrons containing 
charm quarks are vital for interpreting the experimental 
results. We present here preliminary results on the lower 
lying spectra~(with angular momentum $L\le 3$ ) of charmonia, 
$D$ and $D_s$ mesons. The calculations have been carried out on 
two different sets of configurations, provided by the 
BMW-c and QCDSF collaborations, both with 
$N_f =2+1$ dynamical flavours.  

This write-up is organised as follows:
in  section 2 we give details of our computational setup and the 
methods used. Then, we present some preliminary results on the $D_s$ and
charmonium spectra before summarising. 

\section{Computational setup}

The two sets of $N_f=2+1$ configurations provided by BMW-c and QCDSF
both were generated using the tree-level Symanzik-improved gluonic action.  The BMW-c
configurations employ tree-level clover improved Wilson fermions, coupled
to links that have undergone two levels of HEX
smearing~\cite{Capitani:2006ni}.  The QCDSF ensembles use
non-perturbatively improved Wilson fermions with stout links in the
derivative terms (SLiNC
action~\cite{Cundy:2009yy}). 

BMW-c has generated ensembles spanning a range of lattice spacings, from
$a\approx 0.054$~fm to $a\approx 0.092$~fm and pion masses, from
$M_\pi \approx 520$~MeV down to $M_\pi \approx120$~MeV, which is below
the physical point. The volumes range from $32^3\times 64$ to
$64^3\times 144$, and the number of measurements per set for the present
study is $\sim 200$. For more details on the configurations see
reference~\cite{Durr:2010aw}. With a reasonable range of lattice
spacings we can perform a controlled continuum limit extrapolation of
spectral quantities.

The configuration generation of QCDSF is at an earlier stage. A
different approach is taken to varying the sea quark masses: these
are chosen by first finding the $SU(3)_{\rm flavour}$-symmetric
point where flavour singlet mass averages take their physical values.
Subsequently, the individual quark masses are varied while keeping the
singlet quark mass $\bar m_q = (m_u+m_d+m_s)/3$ constant
\cite{Bietenholz:2010jr,Bietenholz:2011qq}. Configurations have been
generated at $\beta=10/g^2=5.5$, corresponding to $a\sim 0.08$~fm, at several
values of $m_{u/d}$ and $m_s$ and two volumes of $24^4\times 48$ and
$32^3\times 64$ lattice points. 

In this study we have so far analysed the two ensembles whose
details are given in table \ref{table1}. The approach taken to varying
the quark mass means the QCDSF ensembles are ideal for studying
flavour symmetry violations, particularly in the $D/D_s$ spectra. The
ensembles chosen are at the symmetric point and the ensemble with the
lightest sea quark mass~(i.e.\ the biggest difference between $m_{u/d}$ and
$m_s$).

\begin{table}[ht!]
\begin{center}
\begin{tabular}{|c|c|c|c|c|}
\hline
$\kappa_l$&  $\kappa_s$ & $a$ fm & \# meas&
 $M_\pi$ (MeV)\\
\hline \hline
$0.12090$&$0.12090$
&  {$0.0795(3)$} & {$941$} & {$442$}\\
$0.12104$&$0.12062$&
{$0.0795(3)$} & {$450$} & {$348$}\\
\hline
\end{tabular}
\caption{\label{table1} Details of the QCDSF configurations used so far in this study.}
\vspace{-5mm}
\end{center}
\end{table}

In order to extract information about the ground and excited states we use the variational
method~\cite{Michael:1985ne,Luscher:1990ck}, with a basis of three
different smearing functions. We choose a sub-set of the bilinear
operators for the mesons given in Ref. \cite{Liao:2002rj}. For each
operator we vary the smearing to enable us to find three smearings
which are sufficiently different, so that the ground state and first
two excited states could be extracted: this corresponds to a smearing
function that is close to optimal for the ground state and two
others that contain significant excited state contributions.  For the
smearing, we used gauge-invariant Wuppertal
smearing~\cite{Gusken:1989ad, Gusken:1989qx} with APE smeared links~\cite{Falcioni:1984ei}.

\vspace{-2mm}
\section{Results}

\begin{figure}[ht!]
\begin{center}
\begin{tabular}{c}
\includegraphics[width=0.87\textwidth]{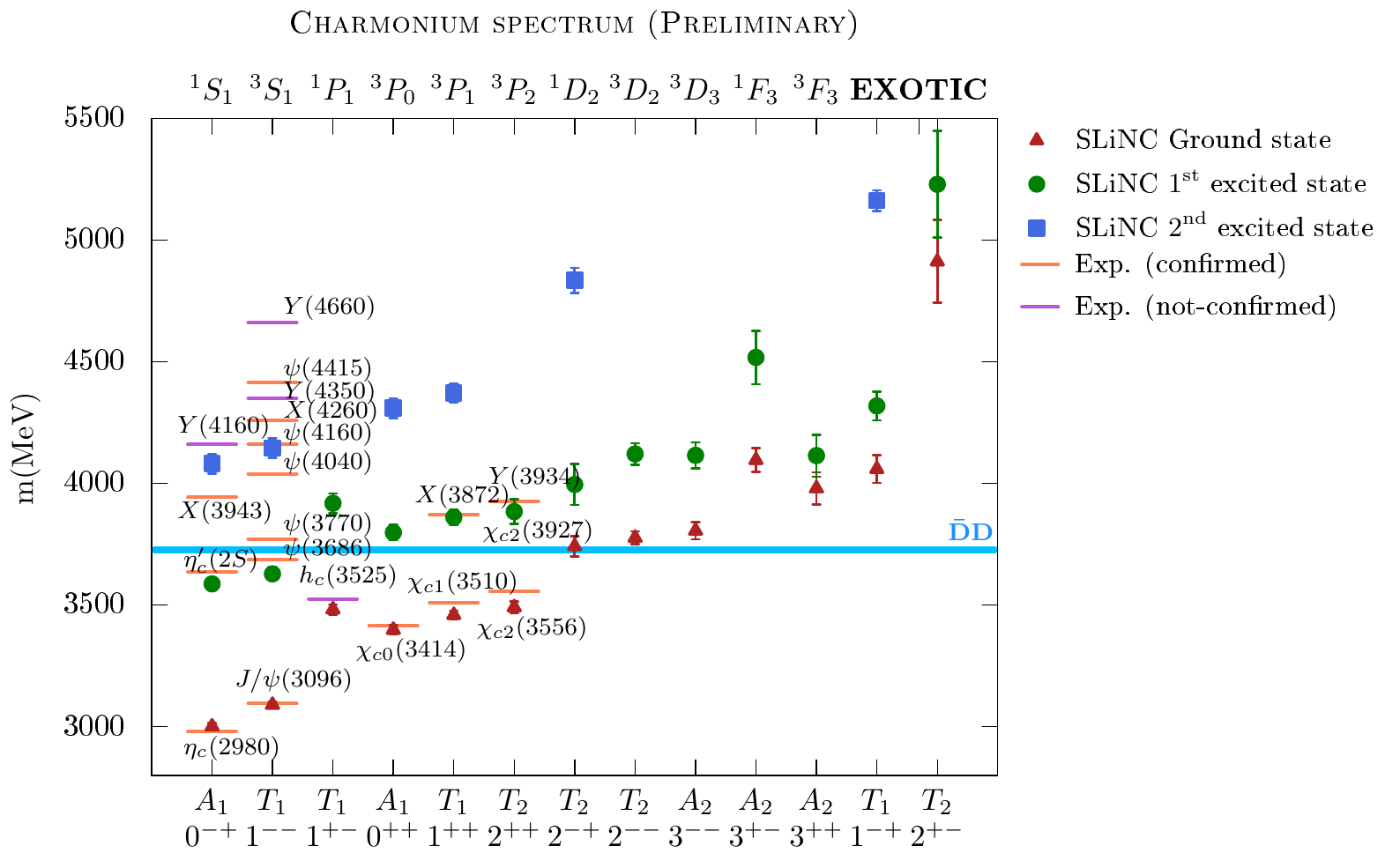} \\
\includegraphics[width=0.87\textwidth]{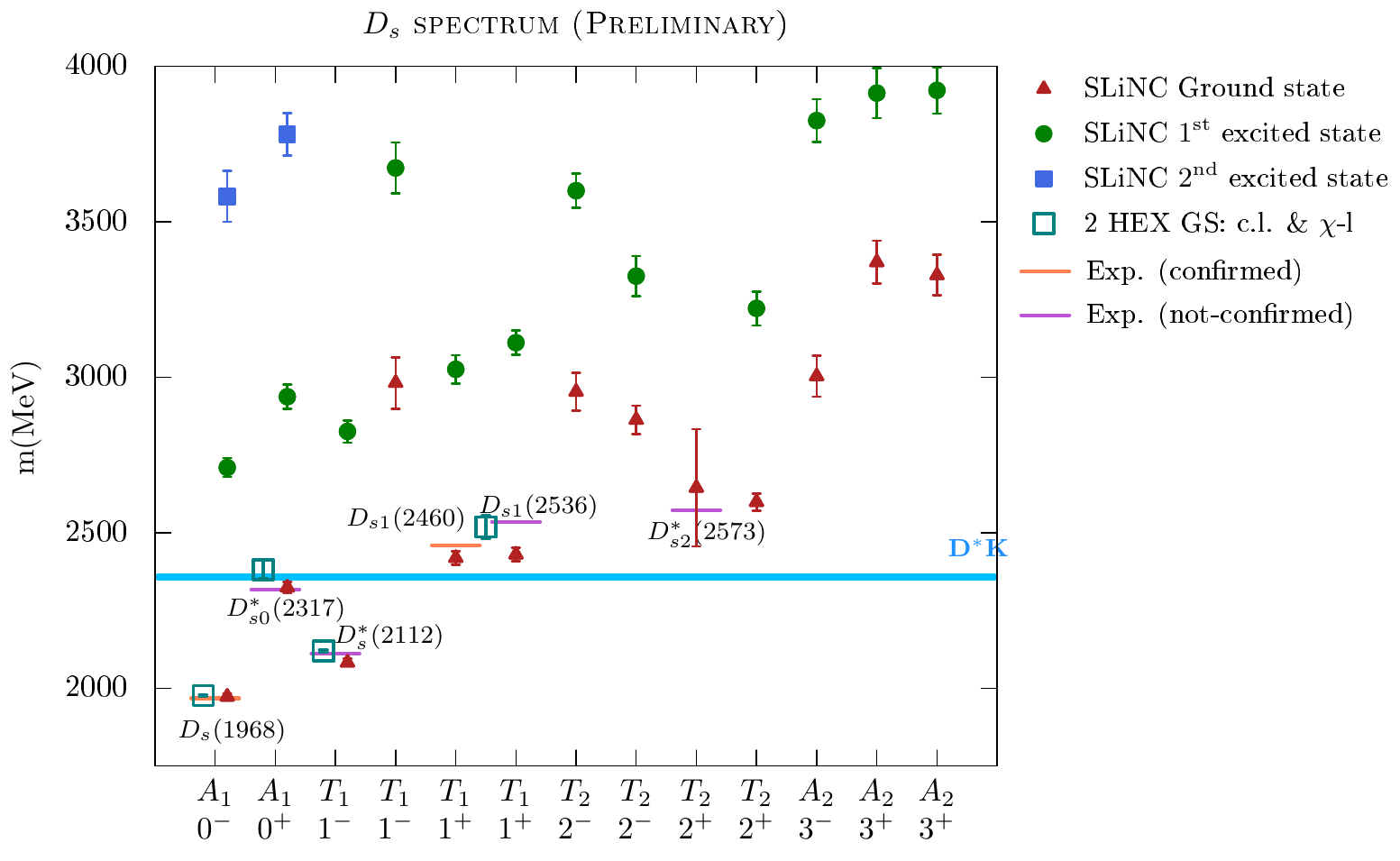} \\ 
\end{tabular}
\parbox{0.8\textwidth}{
\caption{\label{figure1}Charmonium (top) and $D_s$ (bottom) spectra
  from $N_f = 2+1$ configurations. The filled points are results from
  the QCDSF ensemble at the flavour symmetric point.  The empty squares
  correspond to the results for the $D_s$ $S$-wave ground states on the BMW-c
  ensembles extrapolated to the continuum limit and to the physical
  quark masses. Only statistical errors are shown.  }}
\vspace{-6mm}
\end{center}
\end{figure}

Figure \ref{figure1} shows the charmonium (top) and $D_s$ (bottom)
spectra.  The filled points correspond to the results obtained from
the QCDSF configurations at the flavour symmetric point. We have been
able to extract the excited states as well as ground state for all
operators. The second excited state is only included if a clear
plateau was observed in the effective mass plots. This holds for
many of the charmonium states but only for two operators in the $D_s$
spectrum. Since the charm quark mass has not been tuned precisely yet,
we have shifted the results so that the spin average 1S mass agrees
with the experimental data. Qualitatively, the experimental spectra
are reproduced. We use the simplest identification of the J quantum
number with lattice representations, but work is
in progress using the methods in~\cite{Dudek:2007wv}.

\begin{figure}[ht!]
\begin{center}
\vspace{-4mm}
\begin{tabular}{cc}
\includegraphics[height=0.45\textwidth,angle=270]
{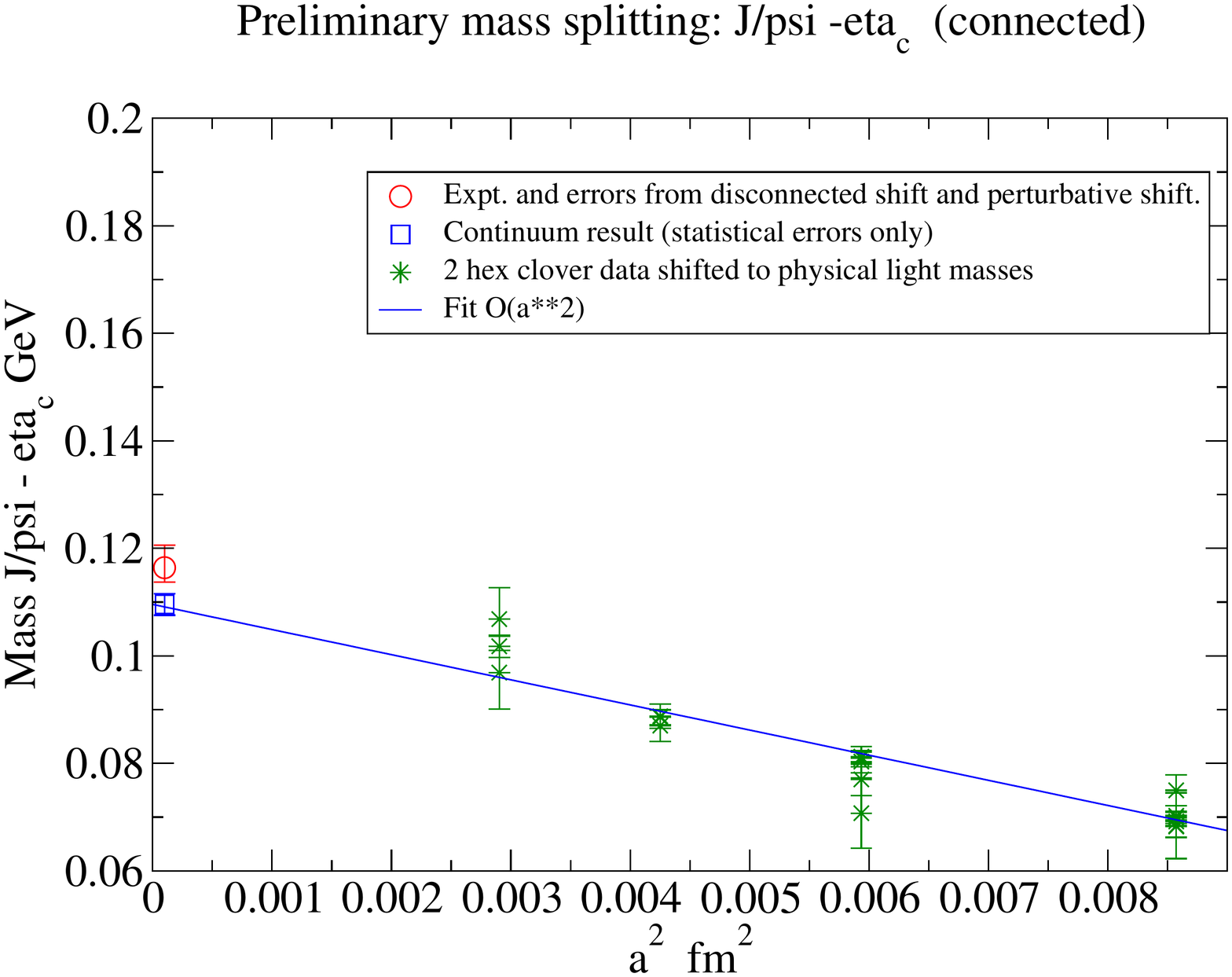} &
\includegraphics[height=0.45\textwidth,angle=270]{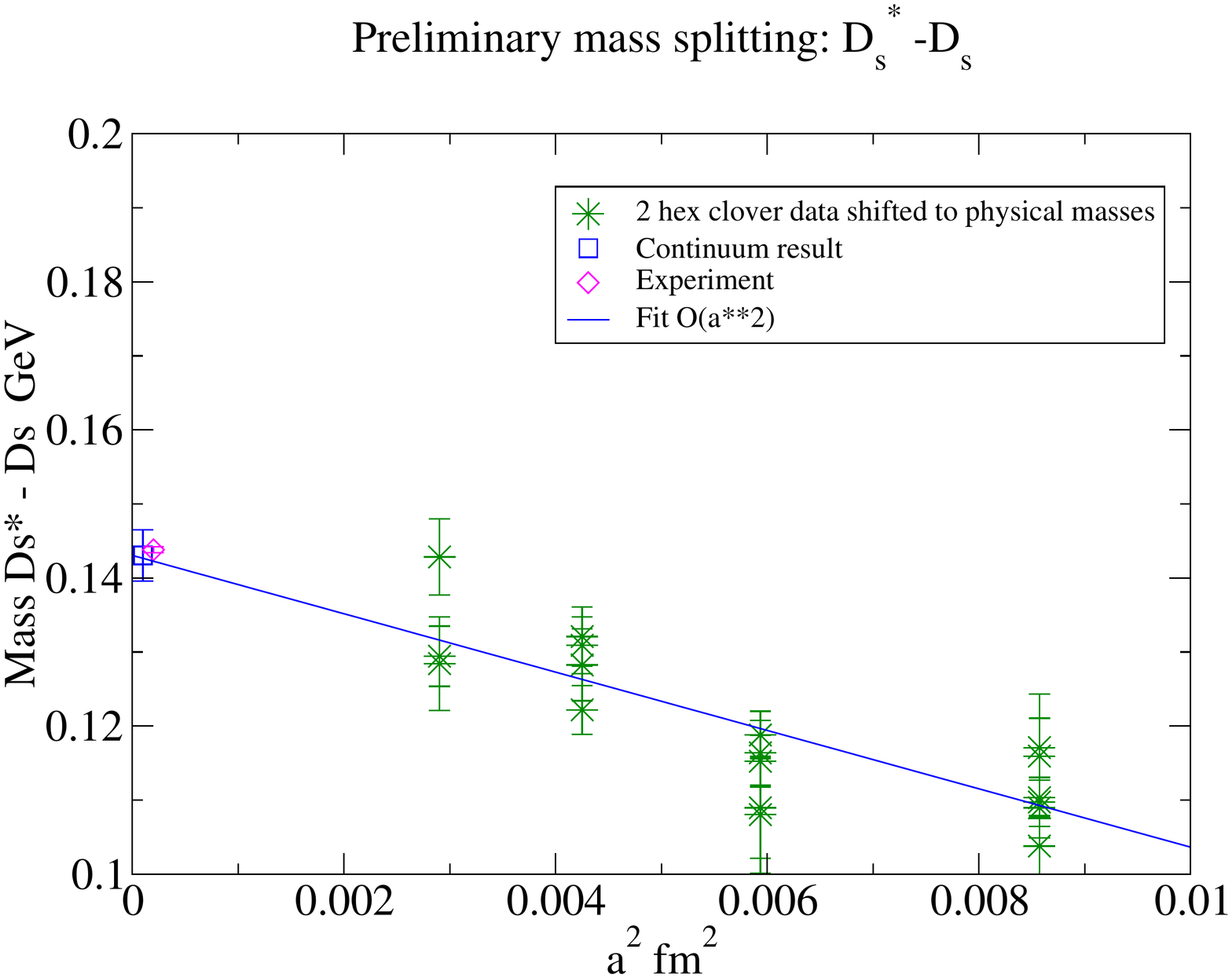} \\ 
\end{tabular}
\parbox{0.8\textwidth}{
\caption{\label{figure2}Charmonium (left) and $D_s$ (right) 
$1S$ hyperfine splittings. Asterisks (green) represent 
the data shifted to the fitted light quark mass values, 
cf. Eq.~(\protect\ref{shift}), the squares (blue) the 
continuum limit extrapolated results. 
}}
\vspace{-4mm}
\end{center}
\end{figure}

In the $D_s$ spectrum plot, we also show results obtained on the BMW-c
configurations that are extrapolated to the physical light quark mass
and continuum limit.  The corresponding extrapolation for the $D_s$
hyperfine splitting is shown in Figure~\ref{figure2}.  
A combined fit encapsulating the lattice spacing and quark mass dependence
is carried out using the fit function,
\be
y^{\rm FIT}(a,M_\pi, m_{\eta_s} |{\bf A}) = (1 +  A_1 a^2) 
\left(1 + A_2 \frac{M^2_{\pi} -\left(  M^{\rm exp}_\pi \right)^2}
{\left(  M^{\rm exp}_\pi \right)^2} + 
A_3\frac{m^2_{\eta_s} - \left( m^{\rm exp}_{\eta_s}\right)^2}
{\left( m^{\rm exp}_{\eta_s}\right)^2} \right),
\ee
where $y$ represents the hyperfine splitting and $m^2_{\eta_s}=2m_{K}^2-m_{\pi}^2$.  Note that, to
demonstrate the quality of the fit, the data points in
figure~\ref{figure2} are shifted to their fitted physical light quark mass
values:
\be\label{shift} 
y^{\rm PLOT}_i = y^{\rm RAW}_i - y_i^{\rm FIT}(a,M_\pi,m_{\eta_s}|{\bf A_{\rm fitted}}) + y^{\rm
  FIT}(a,0,0|{\bf A_{\rm fitted}}).  \ee
This is necessary as we have results at multiple strange quark masses
for some $\beta$s.  There is a significant dependence in the results
on the lattice spacing. However, with values at 4 lattice spacings the
extrapolation is under control and consistency is found with
experiment. Only statistical errors have been considered so far -
a full analysis of the systematic errors is underway.

Similar results are also shown in figure~\ref{figure2} for the
charmonium hyperfine splitting. In this case an asymmetric error bar
has been added to the experimental result to reflect the fact that we have
not included the disconnected contributions. This error bar is $+4$ and $-2.4\,$MeV
based on the results of reference~\cite{Levkova:2010ft}  and the 
estimates of~\cite{Follana:2006rc}, respectively. 
 
\begin{figure}[ht!]
\begin{center}
\includegraphics[width=0.75\textwidth]
{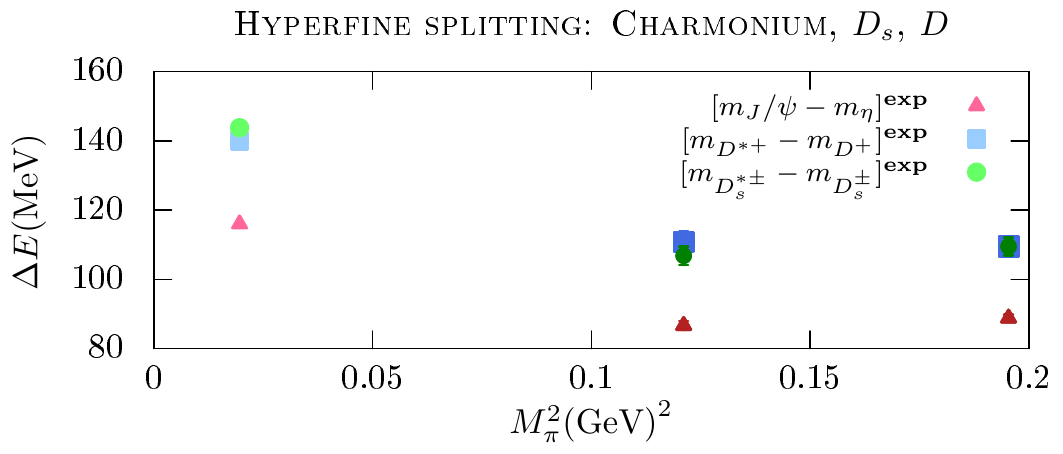}\\ 
\parbox{0.8\textwidth}{
\caption{\label{figure3}Hyperfine splittings (QCDSF ensembles). 
Triangles (dark red) correspond to the charmonium, squares (blue)
to $D$-mesons and circles (green) to $D_s$-mesons. Note that 
for the larger pion mass the latter two data points  coincide since this
corresponds to the flavour symmetric point. }}
\vspace{-4mm}
\end{center}
\end{figure}

Figure~\ref{figure3} shows the hyperfine splitting for the $D$, $D_s$
and charmonium states for the two analysed QCDSF ensembles.
There is little dependence on the pion mass and
the splittings are smaller than the experimental values. However, the
charm quark mass needs to be tuned more precisely and, as observed in
the BMW-c analysis, the continuum and quark mass extrapolation needs
to be performed.

\section{Summary and outlook}

We have presented the current status of an on-going 
project to study hidden and open charm meson spectra
on the lattice. Results have been obtained from two sets 
of configurations with different actions: SLiNC, 
where the singlet quark mass is kept fixed and 2HEX,
where the simulated pion masses range down to 120 MeV.
The results presented include the charmonium and $D_s$ 
spectra and the $1S$ hyperfine splitting. For the latter and 
the ground states of $D_s$ the continuum limit and 
physical mass extrapolations have been performed. 
No other systematics nor disconnected diagrams have
been included as yet. 

In the near future, we expect to perform measurements 
for different pion masses using the QCDSF ensembles 
as well as for different volumes and for two different
lattice spacings so that we can control
the discretization errors, the finite volume effects and
enable an extrapolation to the physical quark masses. 
In addition, we plan to investigate the singly and doubly 
charmed baryons. 
The necessary methods, including all-to-all propagator
techniques~\cite{Bali:2009hu}, have been developed to quantify
mixing in the charmonium sector~\cite{Bali:2009er}
between states of different $L$, with lighter
flavour-singlet states and with $D\overline{D}$ molecular states, 
for $N_f = 2$ QCDSF configurations. We plan to extend this
to the $N_f=2+1$ ensembles.

For the BMW-c configurations, the study of the excited 
states of $D_s$ via the variational method has
started. The systematics of the 
continuum limit and quark mass extrapolations are going
to be 
included.

\section*{Acknowledgements}

The numerical calculations were performed on the SGI Altix ICE
machines at HLRN
(Berlin-Hannover, Germany) and the BlueGene/P (JuGene) and the Nehalem
Cluster (JuRoPA) of the J\"ulich Supercomputer Center. We have made
use of the Chroma software package~\cite{Edwards:2004sx} for some of
the analysis.  This work was supported by the EU ITN STRONGnet, the I3
HadronPhysics2 and the DFG SFB/Transregio 55.  Sara Collins
acknowledges support from the Claussen-Simon-Foundation (Stifterverband
f\"ur die Deutsche Wissenschaft).


\providecommand{\href}[2]{#2}\begingroup\raggedright

\end{document}